\documentclass[conference]{IEEEtran}
\usepackage{psfrag}
	\psfrag{EbNo}[Bc][Bc][0.7]{$10 \log_{10} (E_{\mathrm b}/N_0)\ \rightarrow$}
	\psfrag{FER}[Bc][Bc][0.7]{FER $\rightarrow$}
	\psfrag{Pf}[B][Bc][0.6]{\hspace*{6.5mm}(a)}
	\psfrag{MSSRS}[B][Bl][0.6]{\hspace*{9mm}(b)}
	\psfrag{FER2}[Bc][Bc][0.7]{FER $\rightarrow$}
	\psfrag{FERi2}[Bc][Bc][0.7]{$\leftarrow p$}
	\psfrag{intl}[Bc][Bc][1]{$l$}
	\psfrag{lba}[Bc][Bc][0.8]{(a)}
	\psfrag{lbb}[Bc][Bc][0.8]{(b)}
	\psfrag{lbc}[Bc][Bc][0.8]{(c)}
\ifCLASSINFOpdf
  \usepackage[pdftex]{graphicx}
\else
  \usepackage[dvips]{graphicx}
\fi
\usepackage[cmex10]{amsmath}
\usepackage{amsthm}
\usepackage{amsfonts}
	\newtheorem{theorem}{Theorem}
	\newtheorem{defi}{Definition}
	\newtheorem{lemma}{Lemma}
\usepackage{color}
\usepackage{fancyhdr}

\def\Cond#1{$\boldsymbol{\mathord C\hskip-1.4pt\mathord o\hskip-.6pt\mathord n\hskip-.6pt\mathord d}\mathord .$\,\itshape{#1}\normalfont}
\def\qed{\hspace*{\fill}~\IEEEQED\par\endtrivlist\unskip}
\definecolor{mygray}{rgb}{0.6,0.6,0.6}

\begin{document}
\pagestyle{fancy} 
\fancyhead[RE,LO]{\color{mygray}Accepted for ISIT 2011 --- Hans~Kurzweil, Mathis~Seidl and Johannes~B.~Huber:\\Reduced-Complexity Collaborative Decoding of Interleaved Reed-Solomon and Gabidulin Codes}
\fancyfoot[LE,RO]{\color{mygray}Hans~Kurzweil, Mathis~Seidl and Johannes~B.~Huber, \today}

\title{Reduced-Complexity Collaborative Decoding of Interleaved Reed-Solomon and Gabidulin Codes}

\author{Hans~Kurzweil$^*$, Mathis~Seidl$^\dagger$ and Johannes B. Huber$^\dagger$\\
	$^*$Department Mathematik, Universit\"{a}t Erlangen-N\"{u}rnberg, Erlangen, Germany\\
	$^\dagger$Lehrstuhl f\"{u}r Informations\"{u}bertragung, Universit\"{a}t Erlangen-N\"{u}rnberg, Erlangen, Germany\\
	Email: {\tt kurzweil@mi.uni-erlangen.de, \{seidl, huber\}@lnt.de}}%

\maketitle

\begin{abstract}
An alternative method for collaborative decoding of interleaved Reed-Solomon codes as well as Gabidulin codes for the case of high interleaving degree is proposed. As an example of application, 
simulation results are presented for a concatenated coding scheme using polar codes as inner codes.
\end{abstract}

\IEEEpeerreviewmaketitle

\section{Introduction}
	Reed-Solomon (RS) codes are used in many applications, often implemented in an interleaved form as outer codes in concatenated code designs. 
	By combining and interleaving a number $l \in \mathbb N$ of RS codewords, correction of long error bursts  affecting only a few symbols 
	of the particular underlying codewords can be achieved.\\
	The standard decoding procedure consists of decoding each of the interleaved codewords separately.
	In recent years, methods have been investigated which try to decode the individual codewords no longer independently but in one step, allowing for 
	error correction beyond half the minimum distance $d$. However, in order to decode the maximum possible number of errors $f\!=\!(d\!-\!2)$, the error vectors 
	are required to be linearly independent.\\
	In \cite{Metz} and \cite{Hasla}, a collaborative decoding algorithm for general linear codes based on Gaussian elimination is derived which is able to 
	correct errors up to $\min \{l,d\!-\!2\}$ by solving a reduced system of $l$ linear equations. 
	Therefore, it is applicable only for situations where $l$ can be chosen sufficiently high. 
	Other methods based on multisequence shift-register synthesis (MSSRS) \cite{FengTzeng}, \cite{Boss1} consider the complete system of key equations 
	leading to an increased error correcting radius beyond $l$ (but likewise smaller than $d\!-\!1$) and an improved decoding performance.\\
	While the decoding complexity is of same order $\mathcal{O}(l f^2)$ for both approaches, in case of high interleaving degrees the first method 
	might be preferable from a computational point of view as the Gaussian elimination allows for parallelized computing of rows and columns 
	(and thus for a reduced decoding delay) in contrast to the sequential structure of the shift register synthesis algorithm.\\
	Our considerations are based on the method from \cite{Metz} and \cite{Hasla}. We adopt the results for the special case of RS codes. 
	In contrast to \cite{Metz}, by using a specific code we obtain a unique solution in terms of an error locator polynomial rather than a superset 
	of the error locations. More importantly, only the first part of each syndrome sequence is required for decoding.\\
	Furthermore, we will show that in the case of concatenated codes with high interleaving degree ($l\geq d-2$), the performance degradation compared to MSSRS is small.
\section{Reed-Solomon Codes and Interleaving}
	The authors are aware that the theory of RS codes is widely known. However, since our considerations are based on extended, non-standard RS codes, 
	a short introduction seems to be necessary as well. Although even more general definitions as in \cite{Mac} would be possible, 
	we define a Generalized Reed-Solomon (GRS) code of length $n$ and dimension $k$ over a finite field $\mathbb F$ with 
	$| \mathbb F | = q$ elements as follows:		\vspace*{-1.5mm}
	\begin{defi}[\bf{Reed-Solomon code}]
		Let $\boldsymbol v := (v_1,\ldots , v_n)\in \mathbb F^n$ be a row vector of $n\leq q$ different elements of $\mathbb F$. 
		Let further $\mathcal P_k$ ($k<n$) be the vector space of polynomials over $\mathbb F$ with degree $<k$. Then a Reed-Solomon code 
		$\mathcal{GRS}(q;n,k,\boldsymbol v)$ is defined as the set of evaluations at $\boldsymbol v$
		\vspace*{-0.8mm}\[\Big\{\big(p(v_1), p(v_2) , \ldots , p(v_n)\big) \in \mathbb F^n \ : \ p \in \mathcal P_k \Big\}\]
		\vspace*{-0.8mm}of all the polynomials from $\mathcal P_k$.
	\end{defi}		\vspace*{-1.2mm}
	\begin{defi}
		A Reed-Solomon-Code $\mathcal{GRS}(q;n,k,\boldsymbol v)$ with length $n=q-1$ and
		\[\boldsymbol v = (\alpha^0,\alpha^1,\ldots,\alpha^{q-2})\]
		with $\alpha$ being a primitive element of $\mathbb F$ will be referred to as $\mathcal{RS}(q-1,k)$.\\
		The extended code of length $n=q$ obtained by adding the zero element of $\mathbb F$ to the vector $\boldsymbol v$ of $\mathcal{RS}(q-1,k)$, i.e.
		\[\boldsymbol v = (0,\alpha^0,\alpha^1,\ldots,\alpha^{q-2}),\]
		will be called $\mathcal{RS}^{\ast}(q,k)$.
	\end{defi}		\vspace*{-0.8mm}
	The code $\mathcal{RS}^{\ast}(q,k)$ has one interesting property which has been proved in a more general form in \cite[p. 304]{Mac} and will be the 
	foundation of our following considerations:
	\begin{lemma}\label{Thdual}
		The dual code of $\mathcal{RS}^{\ast}(q,k)$ is $\mathcal{RS}^{\ast}(q,q-k)$, i.e. the Vandermonde matrix
		\begin{equation*}
			\boldsymbol{H} = \begin{pmatrix}
					1 & 1 & 1 & 1 & \ldots & 1\\
					0 & 1 & \alpha & \alpha^2 & \ldots & \alpha^{q-2}\\
					0 & 1 & \alpha^2 & \alpha^4 & \ldots & \alpha^{2(q-2)}\\
					\vdots & \vdots & \vdots & \vdots & \ddots & \vdots \\
					0 & 1 & \alpha^{m-1} & \alpha^{2(m-1)} & \ldots & \alpha^{(q-2)(m-1)}\\
			\end{pmatrix}
		\end{equation*}
		with $m:=n-k$ is a possible parity check matrix of $\mathcal{RS}^{\ast}(q,k)$.
	\end{lemma}
	According to Lemma \ref{Thdual}, syndrome calculations in the case of $\mathcal{RS}^{\ast}$ codes are actually polynomial evaluations. 
	Moreover, we will show in the following that linear combinations of the rows of $\boldsymbol H$ lead directly to coefficients of polynomials, 
	the roots of which will specify the error locations.\\
	By grouping $l \in \mathbb N$ codewords of $\mathcal{GRS}(q;n,k,\boldsymbol v)$ column-wise to a $(n\times l)$-matrix, we obtain a linear code of length $(l\cdot n)$, 
	dimension $(l\cdot k)$ and minimum distance $(n-k+1)$ like the individual codes. 
	\begin{defi}[\bf{Interleaved Reed-Solomon (IRS) code}]
		Given a certain Reed-Solomon code $\mathcal C :=\mathcal{GRS}(q;n,k,\boldsymbol v)$, we define 
		an Interleaved Reed-Solomon code $\mathcal{IRS}(q;l,n,k,\boldsymbol v)$ of interleaving degree $l$ as the set of $(n \times l)$-matrices
		\begin{equation*}
			\bigg\{\boldsymbol{A} = \Big(\boldsymbol{a}^{(1)} , \boldsymbol{a}^{(2)} , \ldots , \boldsymbol{a}^{(l)}\Big) \ : \ 
			(\boldsymbol{a}^{(i)})^{\top} \in \mathcal C \ ,\  i=1,\ldots ,l\bigg\},
		\end{equation*}
		each consisting of $l$ column-wise arranged codewords from $\mathcal C$. In case of $\mathcal C = \mathcal{RS}^{\ast}(q,k)$, the resulting IRS code 
		will be referred to as $\mathcal{IRS}^{\ast}(q,l,k)$.
	\end{defi}
\section{Collaborative Decoding}
	Assume now that a codeword $\boldsymbol{A} \in \mathcal{IRS}^{\ast}(q,l,k)$ is transmitted over an additive noise channel, so that
	\[\boldsymbol{Y} = \boldsymbol{A} + \boldsymbol{E} \ \in \mathbb F^{n\times l}\]
	with some error matrix $\boldsymbol{E}\in \mathbb F^{n\times l}$ is received at the channel output. Let $\boldsymbol{E}$ be a matrix with exactly 
	$f\in \mathbb N$ non-zero rows. We denote $\mathcal F$ the set of indices of these erroneous rows. 
	(Clearly, at first $f$ and $\mathcal F$ are unknown to the decoder.)\\
	For collaborative decoding, we arrange the $l$ syndrome sequences generated from $\boldsymbol Y$ as columns of a so-called 
	syndrome matrix $\boldsymbol{S}$. The computation can be written formally as a matrix multiplication of $\boldsymbol{Y}$ with the parity check matrix 
	$\boldsymbol{H}$ of the underlying $\mathcal{RS}^{\ast}$ code:
	\begin{equation}\label{Synd}
		\boldsymbol{S}=\boldsymbol{H} \cdot \boldsymbol{Y} = \boldsymbol{H} \cdot (\boldsymbol{A}+\boldsymbol{E}) = \boldsymbol{H} \cdot \boldsymbol{E} = 
		\boldsymbol{H}^{\mathcal F} \cdot \boldsymbol{E}_{\mathcal F}
	\end{equation}
	with $\boldsymbol{H}^{\mathcal F}$ and $\boldsymbol{E}_{\mathcal F}$ denoting the submatrices of $\boldsymbol{H}$ and $\boldsymbol{E}$ consisting only of those 
	columns and rows, respectively, whose indices are contained in $\mathcal F$.
	The last equivalence holds because all other rows of $\boldsymbol{E}$ are zero. 
	The syndrome matrix takes the form
	\begin{equation}
		\boldsymbol{S} = \begin{pmatrix}
				s_1^{(1)} &  \ldots & s_1^{(l)}\\
				\vdots &  \ddots & \vdots \\
				s_{n-k}^{(1)} &  \ldots & s_{n-k}^{(l)}\\
			\end{pmatrix} \ \in \mathbb F^{(n-k)\times l}.
	\end{equation}
	Instead of successively solving (for increasing $f^\ast$) the complete system of $l\cdot (n-k-f)$ key equations	\vspace*{-1mm}
	\begin{align}
		s_i^{(m)} = \sum_{j=1}^{f^\ast}\lambda_j s_{i-j}^{(m)},\qquad &i=f^\ast+1,\ldots , n-k \\[-2ex]
									& m=1,\ldots ,l \notag 
	\end{align}
	for the coefficients $\lambda_j \in \mathbb F$ of the error locator polynomial ($\lambda_0 :=1$),
	we use a subsystem consisting of $l$ equations only to be solved:	\vspace*{-2mm}
	\begin{align}\label{keyeq}
		s_{f^\ast+1}^{(m)} = \sum_{j=1}^{f^\ast}\lambda_j s_{f^\ast+1-j}^{(m)}, \qquad & m=1,\ldots ,l
	\end{align}
	From \eqref{keyeq} it follows immediately that the error correcting radius, i.e. the maximum number of erroneous rows that can be corrected, 
	cannot be greater in our case than
	\begin{equation}
		f_{\mathrm{max}} := \min \{l,d\!-\!2\}
	\end{equation}
	The decoding task then plainly consists in determination of the row of $\boldsymbol S$ with smallest index $(f^\ast +1)$ that can be written as a 
	linear combination of the former rows. By applying the Gauss-Jordan algorithm to the columns of $\boldsymbol S$, we obtain the 
	\underline{r}educed \underline{c}olumn \underline{e}chelon \underline{f}orm ($\mathrm{rcef}$) of $\boldsymbol S$:
	\begin{equation}\label{rref}
		\mathrm{rcef}(\boldsymbol S) = \begin{pmatrix}
			1 & \ldots & 0 & 0 & \ldots & 0 \\
			\vdots & \ddots & \vdots & \vdots & \ddots & 0 \\
			0 & \ldots & 1 & 0  & \ldots & 0 \\
			\lambda_1 & \ldots & \lambda_{f^\ast} & 0 & \ldots & 0 \\
			\vdots & \vdots & \vdots & \vdots & \vdots & \vdots
		\end{pmatrix}.
	\end{equation}
	Our decoding algorithm will be successful whenever the following two conditions are fulfilled:\\[1ex]
	\Cond1 : \quad The non-zero rows $\boldsymbol{E}_i$ $(i \in \mathcal F )$ of the error\\
			\hspace*{19mm}matrix are linearly independent.\\
	\Cond2 : \quad $f \leq f_{\mathrm{max}}$ holds.\\[1ex]
	Actually, \Cond1 and \Cond2 are both necessary and sufficient for correct decoding. 
	We will make some remarks on the linear independence condition \Cond1 later in section \ref{probf}.
	\begin{theorem}\label{mainth}
		If \Cond1 and \Cond2 are fulfilled, then $f^\ast = f$ and the polynomial		\vspace*{-1.6mm}
		\[\Lambda (x) = x^{f^\ast} - \sum_{j=1}^{f^\ast}\lambda_{j}x^{j-1} \]
		built from the elements $\lambda_j$ from \eqref{rref} is the error locator polynomial, i.e.		\vspace*{-1mm}
		\[\Lambda (v_i) = 0 \quad \Leftrightarrow \quad i \in \mathcal F\]		\vspace*{-1.5mm}
		holds.  
	\end{theorem}
	\begin{IEEEproof}
		See the Appendix.
	\end{IEEEproof}
	Obviously, only the first $(f+1)$ rows of $\boldsymbol S$ rather than the complete (length $n-k$) syndrome sequences are necessary for finding $\Lambda$. 
	Thus, especially if the actual number of errors is small, the computational complexity can be reduced significantly by successive calculation of the rows of $\boldsymbol S$. 
	We will discuss this version in section \ref{complex}.
\section{Codeword reconstruction}
	Given the (correctly computed) set $\mathcal F$ of erroneous columns of $\boldsymbol{Y}$, we are now able to reconstruct $\boldsymbol{E}_{\mathcal F}$ and therefore
	$\boldsymbol{A}=\boldsymbol{Y}-\boldsymbol{E}$. The matrix equation
	\vspace*{-1.3mm}\begin{equation}\label{reco}
		\boldsymbol{S} = \boldsymbol{H}^{\mathcal F} \cdot \boldsymbol{E}_{\mathcal F}
	\end{equation}
	defines an over-determined system of linear equations consisting of $l \cdot (n-k)$ equations and $l\cdot f$ unknowns. Since we know that \eqref{reco} must have 
	a unique solution and since the first $f$ rows of the (Vandermonde!) matrix $\boldsymbol{H}^{\mathcal F}$ are linearly independent, 
	we can restrict to the smaller system		\vspace*{-1.3mm}
	\begin{equation}\label{reco2}
		\boldsymbol{S}_{[f]} = \boldsymbol{H}_{[f]}^{\mathcal F} \cdot \boldsymbol{E}_{\mathcal F}
	\end{equation}
	with $\boldsymbol{S}_{[f]}$ and $\boldsymbol{H}_{[f]}^{\mathcal F}$ denoting the matrices consisting of the first $f$ rows of 
	$\boldsymbol{S}$ and $\boldsymbol{H}^{\mathcal F}$, respectively. 
	As mentioned before, $\boldsymbol{H}_{[f]}^{\mathcal F}$ is a quadratic - and thus invertible - Vandermonde matrix and \eqref{reco2} actually an interpolation problem.\\
	Note that also for calculation of the error values the last rows $\boldsymbol S_{f+2},\ldots,\boldsymbol S_{n-k}$ of the syndrome matrix $\boldsymbol S$ are not required.
\section{Failure probability}\label{probf}
	As shown before, in case of $f \leq f_{\mathrm{max}}$ the success of the decoding procedure solely depends on the linear independence of the error vectors.
	If we assume that the $\boldsymbol{E}_i$ are random vectors uniformly distributed over $\mathbb F^l \setminus \{\boldsymbol{0}\}$, 
	the probability that \Cond1 is not fulfilled, i.e. that the $\boldsymbol{E}_i$ are linearly dependent, can be overbounded for $f\geq 2$ by
	\begin{equation}
		q^{-(l+1-f)} \cdot \frac{1-q^{-f}}{1-q^{-1}}\approx q^{-(l+1-f)},
	\end{equation}
	as shown in \cite{Metz}. Clearly, the decoder certainly fails if the number of erroneous columns exceeds $f_{\mathrm{max}}$. Thus,
	\begin{equation}\label{PfIRS}
		P_f(f,l) \leq \begin{cases} 0 & f<2\\q^{-(l+1-f)} & 2 \leq f\leq f_{\mathrm{max}}\\ 1 & \mathrm{else}\end{cases}
	\end{equation}
	holds as an upper bound for the failure probability under assumption of uniformly distributed 
	error vectors.\\
	Compared to the failure probabilities of the Feng-Tzeng algorithm as derived in \cite{Boss1}, the probabilities for the decoding to fail are equivalent for 
	$f\!=\!f_{\mathrm{max}}\!=\!(n-k-1)$, but decline significantly slower as $f$ decreases. Moreover, for $l<(n-k-1)$ the error correction radius 
	of the Feng-Tzeng algorithm, i.e. the number of correctable errors, is in general strictly greater than in our case.\\
	In concatenated code designs where the columns of an outer IRS code are encoded by an inner block code, the overall frame error rate (FER)
	can be analytically determined by		\vspace*{-1.5mm}
	\begin{equation}
		\mathrm{FER} = \sum_{t=2}^{N}\binom N t \cdot P_f(t,l) \cdot p^t \cdot (1-p)^{N-t}.
	\end{equation}
	with $p$ being the frame error rate of the inner code.\\
	\begin{figure}[ht]
		\begin{center}
			\includegraphics[width=0.47\textwidth]{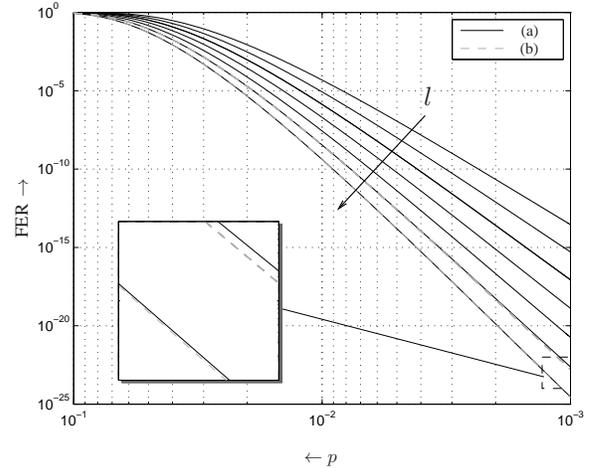}
		\end{center}
		\caption{Bounds on the frame error rate (FER) for a $(204,188)$ IRS code with $l\!=\!9\ldots 15$ for collaborative decoding using 
			(a) Gaussian elimination and (b) multisequence shift-register synthesis (MSSRS)}\label{fig_bounds}
	\end{figure}
	Fig. \ref{fig_bounds} depicts the bounds on the FER as a function of the inner code error rate $p$ for various interleaving degrees in the range 
	from $9$ to $15$. Here, a $(204,188)$ shortened RS code like in the DVB standard \cite{DVB} is used. 
	For comparison, the failure bounds for MSSRS are plotted as dashed grey lines.\\
	Whereas for small interleaving degrees $l$ the multisequence decoder (the lines of which coincide for $l=9\ldots 14$ !) clearly outperforms our method, 
	for $l\geq 14$ the performance of both approaches is nearly identical.
\section{Complexity}\label{complex}
	The computational complexity of the Gauss-Jordan algorithm is of same order $\mathcal O(lf^2)$ like the independent decoder as well as the 
	multisequence synthesis algorithm (with $l$ and $f$ as defined before).\\
	As already mentioned, only the first $(f+1)$ rows rather than the whole matrix $\boldsymbol{S}$ are actually necessary for determination 
	of the error locator polynomial. 
	In particular, when $f\ll l$ the computational complexity may be significantly reduced by combining syndrome calculation and actual decoding, 
	i.e. by applying a decoding algorithm which in each step $t=1\ldots f+1$ calculates one additional row of $\boldsymbol S$ and performs 
	Gaussian elimination on the corresponding $(t \times l)$-submatrix of $\boldsymbol S$ until a linearly dependent row is detected. 
	Therefore, the complexity of the syndrome calculations reduces to $\mathcal O(lnf)$ rather than $\mathcal O(lnd)$. 
	In the extreme case that no errors have occurred, the decoder stops without performing even a single finite field multiplication since, due to the 
	special form of $\boldsymbol H$, the first coefficient of the $l$ syndrome sequences each simply consists of a column sum of $\boldsymbol Y$.\\
	Moreover, in each step of the Gaussian elimination, the $l$ columns of the syndrome matrix $\boldsymbol{S}$ may be transformed at the 
	same time by a parallel implementation of the decoder in order to achieve further reductions in decoding delay.
\section{A Note on Gabidulin codes}
	Gabidulin codes \cite{Gabi} are a class of linear rank metric codes which play an important role in random linear network coding \cite{RNC}. 
	Their codewords can be represented either as $(m \times n)$-matrices over a finite field with $q$ elements $\mathbb F_q$ 
	or equivalently as vectors over the extension field $\mathbb F_{q^m}$.\\
	For decoding $l$-interleaved Gabidulin codes, a key equation for computing the error span polynomial from the syndromes can be derived 
	(cf. \cite{IGabi}) analogously to the case of IRS codes \eqref{keyeq}:		\vspace*{-1.7mm}
	\begin{equation}\label{keyeqGabi}
		\boldsymbol S_{f+1} = \sum_{j=1}^{f}\lambda_j \big( \boldsymbol S_{f+1-j} \big)^{[j]} \ \in \mathbb F_{q^m}^l,
	\end{equation}
	where the operator $[j]$ denotes the pointwise applied $q^j$-th power of a vector.\\
	With a slight modification, i.e. by raising the current row $\boldsymbol S_j$ of $\boldsymbol S$ to the power $[j]$ in each elimination step, 
	the algorithm presented in this paper is applicable to decoding of interleaved Gabidulin codes as well. 
	In this case, an analogue to the failure bound for IRS codes \eqref{PfIRS} is obtained for decoding interleaved Gabidulin codes:
	\begin{theorem}
		The failure probability in case of uniformly distributed rank-$f$ error words is upper bounded by
		\begin{equation}
			P_f^{\mathcal G}(f,l) \leq \begin{cases} 0 & f<2\\4 \cdot (q^m)^{-(l+1-f)} & 2 \leq f\leq \min \{l,d\!-\!2\}\\ 1 & \mathrm{else}\end{cases}
		\end{equation}
		where $d$ denotes the minimum rank distance of the underlying Gabidulin code.
	\end{theorem}
	\begin{IEEEproof}
		A proof similar to the proof of Theorem 3.11 in \cite{OverDiss} is given in the Appendix.
	\end{IEEEproof}
\section{Code Concatenation}
	The derivations of failure bounds on the FER here as well as in \cite{Boss1} remain valid only as long as the error vectors are 
	distributed uniformly over $\mathbb F^l \setminus \{\boldsymbol{0}\}$. 
	In \cite{Boss1}, it was demonstrated for small interleaving degrees ($l=3$) that the performance degradation due to a different error distribution 
	can be neglected when using tailbiting convolutional codes as inner codes.
	Unfortunately, in case of large inner code lengths ($>100$) this result does not hold anymore as convolutional codes produce error vectors of relatively small weight.
	Instead of applying randomizing methods which do not only permute the error bits but also increase their amount while introducing additional computational complexity, 
	we propose the use of an alternative inner coding.\\
	Polar codes \cite{Polar}, first introduced by E. Ar{\i}kan, are decoded by a low-complexity successive decoder which generates estimations 
	on the source bits one after another, each depending on the decisions made before. In case of a wrong decision, long error sequences up to the 
	end of the codeword are produced. 
	This fact (which could usually be seen as a drawback) makes polar codes well suited as 
	inner codes in our case. However, the polar successive decoder happens to fail at certain bits significantly more likely than at other ones. 
	Therefore, in order to meet \Cond1 it appears favorable to apply random permutations to the information bits of the polar code, 
	different for each row of the IRS codeword.
\section{Simulation results}
	Finally, we present some examplary simulations to demonstrate the tightness of the derived bounds and to show that the assumption of 
	random error vectors can be realized in practice.
	As an application example closely related to the DVB-X standards (first version) \cite{DVB}, a (256,128) polar code as inner code 
	together with the (204,188) IRS code from section \ref{probf} is used. 
	As the corresponding IRS code is able to correct up to $15$ erroneous columns, we chose an interleaving degree of 
	$l\!=\!16$ rather than $l\!=\!12$ in the DVB standard. 
	\begin{figure}
		\begin{center}
			\includegraphics[width=0.40\textwidth]{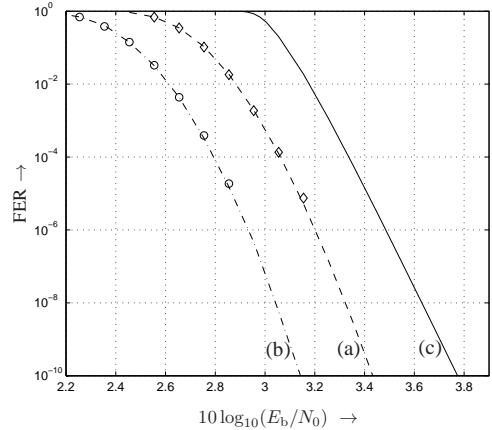}
		\end{center}
		\caption{Simulated and analytical frame error rate (FER-) performance of a $l\!=\!16$ concatenated code with an outer RS(204,188) code and 
			(a) an inner (256,128) polar code with standard decoding, (b) an inner (256,128) polar code with improved 
			decoding, (c) an inner rate 1/2 convolutional code with constraint length $K\!=\!7$}\label{fig_sim2}
	\end{figure}
	In Fig. \ref{fig_sim2}, the frame error rate (FER-) performance of the concatenation design including polar codes is compared to a concatenation of an 
	inner rate 1/2 convolutional code (constraint length $K \! = \! 7$) with an independently 
	decoded outer $(204,188)$ RS code as used in the DVB standard. By using an improved polar decoding scheme as considered in \cite{Seidl}, 
	the performance can be further enhanced by about $0.3 \, \mathrm{dB}$. Simulation results of those improved polar codes are represented 
	by circular markers. 
	Fig. \ref{fig_sim2} shows that the simulation points of both decoding schemes meet the theoretical failure bound very accurately.
	Both IRS-polar concatenation schemes clearly outperform the DVB code in terms of frame error rates as well as of computational complexity.	\vspace*{-1mm}
\appendix
\subsection{Proof of Theorem 1:}
		Let $f < (n-k)$ denote the actual number of erroneous rows and $\mathcal F$ the set of their indices 
		(both being at first unknown to the decoder). 
		The square submatrix of $\boldsymbol{H}$		\vspace*{-1mm}
		\begin{equation}
			\boldsymbol{K} := \big( \boldsymbol{H}^{\mathcal F} \big)_{[f]} \in \mathbb F^{f\times f},
		\end{equation}
		consisting of the first $f$ rows and the columns with indices from $\mathcal F$ of $\boldsymbol{H}$, is a Vandermonde and thus 
		non-singular matrix of full rank $f$. Therefore, the $(f+1)$th row of $\boldsymbol{H}^{\mathcal F}$
		\begin{equation}
			\boldsymbol{\mu} := \big( \boldsymbol{H}^{\mathcal F} \big)_{f+1} \in \mathbb F^{f}
		\end{equation}
		is a unique linear combination of the first $f$ rows $\boldsymbol{K}_j$ of $\boldsymbol{K}$:		\vspace*{-1.1mm}
		\begin{equation}\label{lincomb}
			\boldsymbol{\mu} = \sum_{j=1}^{f}\lambda_j \boldsymbol{K}_j
		\end{equation}
		for some $\lambda_j \in \mathbb F$, $j=1,\ldots , f$. 
		We will now demonstrate how these coefficients $\lambda_j$ can be derived from the syndrome matrix $\boldsymbol S$.
		Let $\varphi$ be the linear mapping defined by $\boldsymbol{E}_{\mathcal F} \in \mathbb F^{f\times l}$:
		\begin{equation}
			\varphi \ : \ \mathbb F^{f} \mapsto \mathbb F^{l} \quad , \quad \boldsymbol v \mapsto \boldsymbol v \cdot \boldsymbol{E}_{\mathcal F}.
		\end{equation}			\vspace*{-0.8mm}
		By definition of $\boldsymbol K$ and $\boldsymbol \mu$,
		\begin{align*}
			\varphi (\boldsymbol{K}_j) &= \boldsymbol{S}_j \quad , \quad j=1,\ldots ,f\\
			\varphi (\boldsymbol \mu)  &= \boldsymbol{S}_{f+1}
		\end{align*}
		By our assumption \Cond1, $\boldsymbol{E}_{\mathcal F}$ is a matrix of rank $f$. Thus, $\varphi$ is an injective mapping, and the vectors 
		$\boldsymbol{S}_j$ $(j=1,\ldots ,f)$ form a basis of the image $\varphi(\mathbb F^f) \subset  \mathbb F^l$ of $\varphi$.\\
		Consequently, the $(f+1)$th row of $\boldsymbol{S}$ is a linear combination of the former, uniquely determined by the very same coefficients 
		$\lambda_j$ as in \eqref{lincomb}:
		\begin{equation}
			\boldsymbol{S}_{f+1} = \varphi(\boldsymbol{\mu}) = \sum_{j=1}^{f}\lambda_j \varphi(\boldsymbol{K}_j) = \sum_{j=1}^{f}\lambda_j \boldsymbol{S}_j.
		\end{equation}
		Thus, the actual number of errors $f$ is given by the row of $\boldsymbol S$ with minimum index that can be written as a 
		linear combination of the preceding rows. It is clear that this index as well as the coefficients $\lambda_j$ can be calculated by performing 
		elementary column operations on $\boldsymbol S$.\\
		Given these coefficients, we define a polynomial		\vspace*{-1mm}
		\begin{equation}
			\Lambda(x) := x^f - \sum_{j=1}^{f}\lambda_j x^{j-1}.
		\end{equation}
		Due to the special form of $\boldsymbol{H}$ (cf. Lemma \ref{Thdual}), the $i$th column $\boldsymbol{H}^i$ consists of the consecutive powers 
		of $v_i\!\in\!\mathbb F$. Consequently,		\vspace*{-2mm}
		\begin{equation}
			0 = \Lambda(x_i) = x_{i}^{f} - \sum_{j=1}^{f} \lambda_j x_{i}^{j-1}
		\end{equation}
		holds if and only if $i \in \mathcal F$.
		Since $\Lambda$ is a polynomial of degree $f=|\mathcal{F}|$, these are obviously the only roots. Therefore,
		$\Lambda$ is the error locator polynomial. \qed
\subsection{Proof of Theorem 2:}
		In the following, we will denote the rank of a $(n \times k)$-matrix $\boldsymbol S$ over an extension field 
		$\mathbb F_{q^m}$ of $\mathbb F_{q}$ by $\mathrm{rank}_{q^m}(\boldsymbol S)$ while the rank of the corresponding 
		$(n \times km)$-matrix over $\mathbb F_{q}$ will be referred to as $\mathrm{rank}_{q}(\boldsymbol S)$.\\
		Let $\boldsymbol E \in \mathbb F_{q^m}^{n \times l}$ be an arbitrary additive error word of an $l$-interleaved Gabidulin code, 
		chosen at random from the set of matrices with $\mathrm{rank}_{q}(\boldsymbol E)=f<f_{\mathrm{max}}$.
		Let further $\boldsymbol S \in \mathbb F_{q^m}^{f \times l}$ denote the submatrix consisting of the first $f$ rows of the corresponding 
		syndrome matrix $\boldsymbol {HE}$. 
		As the parity check matrix $\boldsymbol H$ is of maximum rank over $\mathbb F_q$, the possible matrices 
		$\boldsymbol S$ are uniformly distributed over the set
		\[\mathcal S_f := \{ \boldsymbol A \in \mathbb F_{q^m}^{f \times l} \ : \ \mathrm{rank}_{q}(\boldsymbol A)=f\}.\]
		It is known (cf. \cite[p. 50]{Lidl}) that the mapping
		\[\sigma_i :\quad  \mathbb F_{q^m} \mapsto \mathbb F_{q^m} \quad , \quad \alpha \mapsto \alpha^{[i]} \qquad (i \in \mathbb N)\]
		defines an automorphism of the field $\mathbb F_{q^m}$. Thus, there exists a one-to-one correspondence $\psi$ between $\mathcal S_f$ 
		and the set of matrices defined by the key equation \eqref{keyeqGabi}:
		\[\psi : \boldsymbol S = \Big( \boldsymbol S_1 , \boldsymbol S_2 , \ldots , \boldsymbol S_f \Big)^\mathrm{T} 
			\mapsto \Big( \boldsymbol S_1 , \boldsymbol S_2^{[1]} , \ldots , \boldsymbol S_f^{[f-1]} \Big)^\mathrm{T}\]
		Obviously, the decoding will only fail if $\mathrm{rank}_{q^m}\big(\psi(\boldsymbol S)\big)\! < \! f$ for $\boldsymbol S \in \mathcal S_f$.
		In this case there exists a nontrivial linear combination $\boldsymbol 0 \neq \boldsymbol v \in \mathbb F_{q^m}^{f}$ such that
		\begin{equation}\label{Null}
			\boldsymbol h \cdot \psi (\boldsymbol S) = \boldsymbol 0 \in \mathbb F_{q^m}^l
		\end{equation}
		holds. Because of $\mathrm{rank}_{q^m}(\boldsymbol h) = 1$ there are at most
		\[N_{f}:= (q^m)^{l(f-1)}\]
		possibilities to choose a matrix $\boldsymbol S \in \mathbb F_{q^m}^{f \times l}$ such that \eqref{Null} is fulfilled.
		On the other hand, it is shown in \cite{OverDiss} that
		\[| \mathcal S_f | \geq \frac{1}{4} \cdot (q^m)^{lf}.\]
		Consequently, for an arbitrary chosen $\boldsymbol v$, the probability that $\mathrm{rank}_{q^m}\big(\psi(\boldsymbol S)\big) < f$ 
		for a randomly chosen matrix $\boldsymbol S$ cannot be greater than
		\[P_{\mathrm{max}} := \frac{N_f}{| \mathcal S_f |} \leq 4 \cdot (q^m)^{-l}.\]
		Now the overall failure probability can be upper bounded by summing up over the number of all distinct null spaces defined by different choices of $\boldsymbol v$. 
		This number is certainly smaller than
		\[\frac{(q^m)^f-1}{q^m-1} \approx (q^m)^{f-1}=:N\]
		because $\boldsymbol v \neq \boldsymbol 0$ and because $\boldsymbol v$ and $\alpha \boldsymbol v$ lead to the same null space for any 
		nonzero $\alpha \in \mathbb F_{q^m}$.\\[0.6em]
		Finally, $P_{f}^{\mathcal G}(f,l)$ is bounded by
		\[P_{f}^{\mathcal G}(f,l) \leq P_{\mathrm{max}} \cdot N \leq 4 \cdot (q^m)^{-(l+1-f)}.\]
		\qed

\bibliographystyle{IEEEtran}		\vspace*{4mm}
\bibliography{IRS}

\end{document}